\documentclass[%
 reprint,
nofootinbib,
 amsmath,amssymb,
 aps,
]{revtex4-2}
\usepackage{CJK}
\usepackage{graphicx}
\usepackage{dcolumn}
\usepackage{bm}
\usepackage{subcaption}
\usepackage{float}

\begin{document}
\begin{CJK*}{UTF8}{gbsn}
\title{Exact work distribution and Jarzynski's equality of a relativistic particle in an expanding piston}
\author{Xianghang Zhang (张湘杭)}
\affiliation{
 School of Physics, Peking University, Beijing 100871, China
}
\affiliation{
 Fakult\"at f\"ur Physik, Ludwig-Maximilians-Universit\"at, M\"unchen 80333, Germany
}

\author{Tingzhang Shi (石霆章)}
\affiliation{
 School of Physics, Peking University, Beijing 100871, China
}

\author{H. T. Quan (全海涛)}\email{htquan@pku.edu.cn}
\affiliation{
 School of Physics, Peking University, Beijing 100871, China
}
\affiliation{
 Collaborative Innovation Center of Quantum Matter, Beijing 100871, China
}
\affiliation{Frontiers Science Center for Nano-optoelectronics, Peking University, Beijing 100871, China}

\date{\today}

\begin{abstract}
We study the nonequilibrium work in a pedagogical model of relativistic ideal gas. We obtain the exact work distribution and verify Jarzynski's equality. In the nonrelativistic limit, our results recover the nonrelativistic results of Lua and Grosberg [J. Phys. Chem. B, \textbf{109}, 6805(2005)]. We also find that, unlike the nonrelativistic case, the work distribution no longer has zeros and the number of collisions in this relativistic gas model is finite.
In addition, based on an analysis of the experimental parameters, we conclude that it is difficult to detect the relativistic effects of the work distribution of the ideal gas in a piston system with the current experimental techniques.
\end{abstract}

\maketitle
\end{CJK*}

\section{\label{sec:introduction}Introduction}
Jarzynski's equality (JE) \cite{jar97} is one of the most elegant results in the field of nonequilibrium statistical physics.
While it is a direct result from Liouville's theorem and is supposed to hold in arbitrary Hamiltonian systems including relativistic systems, the work distribution of a relativistic system under an arbitrary work protocol has not been explored so far.
In this article we study the work distribution of a simple relativistic model consisting of only a one-dimensional cylinder, a piston, and relativistic ideal gas.
Previous work on this model has been done and the JE verified, but in a very limited range of parameters \cite{nol09} such that the collision only happens no more than once for all particles.
Moreover, we still lack detailed information of the work distribution of the system giving rise to various fluctuation theorems, which plays a fundamental role in nonequilibrium statistical physics.
A Newtonian approximation of our model \cite{lua05}, however, has yielded analytic results for both the JE and the work distribution, suggesting the solvability in the relativistic regime.
Relevant to the nonrelativistic ideal gas in a piston, the nonequilibrium work distribution of quantum gas in an expanding piston \cite{Quan2012PRE, Gong2014PRE} has been studied and the JE verified.
It is desirable to extend those studies to the relativistic regime.
In this article, we analytically compute the work distribution and verify the JE in this simple setup.
Our model can be viewed as a relativistic generalization of the one in Ref.~\cite{lua05} and recovers it in the low-speed and low-temperature limit.
Such a toy model would be helpful for understanding more complicated and more realistic systems' behavior under relativistic conditions.
As we will see in this paper, although the work distribution changes drastically in the extreme relativistic regime, the equality itself is independent of the microscopic dynamics.
In the following, we will study the nonequilibrium work distribution of relativistic gas in an expanding piston.
The results can serve as a pedagogical example and provide intuitive insights into the robustness of the JE.
Also, we will show that it is difficult to detect the relativistic effect of the work distribution with the current experimental techniques.

The article is organized as follows:.
In Sec.~\ref{sec:model} we analytically calculate the work distribution of a relativistic particle in an expanding piston system and verify the JE.
In the nonrelativistic limit, we recover the results in Ref.~\cite{lua05}.
In Sec.~\ref{sec:dis} we discuss the main results and summarize our work.

\section{\label{sec:model}Relativistic Piston Model}
The model we consider here is nothing more than some ideal gas consisting of $N$ molecules inside a one-dimensional cylinder.
Suppose initially the length of the vessel is $L$ and the gas is initially at the inverse temperature $\beta$, where $\beta = 1/k_{B} T$, with $k_B$ the Boltzmann constant and $T$ the temperature.
We now expand the piston outward at the speed $v_p$ and stop it after a time interval $\tau$.

Under the assumption that the gas is ideal, all particles contribute to both the work $W_\tau$ done by the system up to time $\tau$ and the difference between the final and the initial free energy $\Delta F$ independently.
Consequently, the JE can be rewritten as
\begin{equation}
    (\langle e^{W_1/k_B T}\rangle)^N = (e^{-\Delta F_1/k_B T})^N,
\end{equation}
where $W_1$ and $\Delta F_1$ denote the work and the change of the free energy per particle, respectively.
We can now see that for the ideal gas, the single-particle quantities $W_1$ and $\Delta F_1$ also satisfy the JE.
Thus, we limit our ideal gas to just a single particle, and from now on we omit the subscript $1$ for the discussion of the single-particle quantities.

\subsection{Trajectory of a single particle}

To calculate the trajectory work as a function of the initial state, we must first determine the trajectory of a single particle.

Let us denote the velocity of the particle after the $n$th collision with the moving piston by $v_{n}.$
Following the moving of a particle, we find out that after the $(n+1)$th collision with the moving piston, the speed of the particle is reduced to
\begin{equation}
    v_{n+1} = \frac{(c^{2} + v_p^2)v_n - 2v_pc^{2}}{c^{2}+v_p^2-2v_p v_n}.
\end{equation}
This result can be derived simply by performing Lorentz transformation twice, noting that the piston is an inertial reference frame during the whole process and the collisions are elastic collisions.
When $v_n, v_p\ll c, v_{n+1}= \left(v_n-2v_P+v_n v_p^2/c^2\right)/\left(1+v_p^2/c^2-2v_n v_p/c^2\right)\approx v_n-2v_p$, which is the result in the nonrelativistic limit.

The solution to the recurrence relation of the particle's speed $v_n$ after the $n$th collision can also be derived as
\begin{equation}\label{eqn:vn}
    v_n = \frac{(c + v)\alpha_p^{2n} - c + v}{(c + v)\alpha_p^{2n} + c - v} c,
\end{equation}
where $v$ is the initial speed of the particle and $\alpha_p$ is a parameter pertaining to the velocity $v_p$ of the moving piston,
\begin{equation}
    \alpha_p = \frac{c - v_p}{c + v_p}.
\end{equation}

Another thing we must figure out is the time $t_{n}$ when the $n$th collision with the moving piston takes place.
We have the recurrence relation

\begin{equation}
    t_{n + 1} = \frac{2L + (v_p + v_n)t_n}{v_n - v_p},
\end{equation}
from which we can derive the expression of $t_n$, 
\begin{equation}\label{eqn:recursionresult}
\begin{split}
    t_n = \frac{2L}{v_{n - 1} - v_p}\left(1+\sum_{i=2}^{n-1}\prod_{j=i}^{n-1}\frac{v_j+v_p}{v_{j-1}-v_p}\right) \\
    + \left(\prod_{j=1}^{n-1}\frac{v_j+v_p}{v_{j-1}-v_p}\right)\cdot\frac{(L \pm x)}{v_{n-1}-v_p}.
\end{split}
\end{equation}
Here $x$ denotes the initial position of the particle.
The sign of $x$ is positive if the initial velocity is towards the moving piston and negative if the initial velocity is away from the moving piston.
So from now on we simply extend the range of $x$ from $[0, L]$ to $[-L, L]$ to remove the negative sign (for details, see Appendix~\ref{app:cov}).
With the expression of $v_n$, the product of a sequence can be simplified as
\begin{equation}
    \prod_{j=1}^n \frac{v_j+v_p}{v_{j-1}-v_p} = 2 \alpha_p^n \left[1 + \alpha_p^{2n}+(-1 + \alpha_p^{2n})\frac{v}{c}\right].
\end{equation}
Finally, we have
    \begin{equation}\label{eqn:tn}
    \begin{split}
         t_n = &\left[(-\alpha_p^{2n}-\alpha_p+\alpha_p^{n+1}+\alpha_p^{n})\frac{v}{c}\right. \\
         &\left.+(-\alpha_p^{n+1}+\alpha_p^{n})\frac{x}{L}+(\alpha_p-\alpha_p^{2n})\right] \\
         &\times\left[-\alpha_p+\alpha_p^{2n}+(\alpha_p+\alpha_p^{2n})\frac{v}{c}\right]^{-1}\cdot\frac{L(1+\alpha_p)}{c(1-\alpha_p)},
    \end{split}
    \end{equation}
which is the time of the $n$th collision between the particle and the moving piston.

The speed of the particle cannot be larger than $c$.
Therefore, during a finite period $\tau$, $n$ cannot take an arbitrarily large number, and the $n$th collision is guaranteed if and only if both
\begin{equation}
        t_n \leq \tau\,\mathrm{and}\, v_{n-1} > v_p
\end{equation}
are fulfilled.

The first requirement ensures that the collision happens before the ending time $\tau$ and the second that the particle can catch up with the moving piston.
The particles with the initial velocity $v \to c$ and initial position $x=L$ undergo the maximum number of collisions.
With these conditions and letting $t_n\leq \tau$ in Eq.~(\ref{eqn:tn}), we find the maximum number of collisions
\begin{equation}
    n \leq N = \left[-\frac{\log\left(1 + \frac{v_p\tau}{L}\right)}{\log\alpha_p}\right] + 1,
\end{equation}
where $[\cdots]$ denotes the integer part of $\cdots.$

Having obtained the number of collisions of every trajectory, we are able to calculate the trajectory work which is a functional of the trajectory and can be determined by the difference of the initial and the final energy of the system.

\subsection{\label{subsec:verification} Direct verification of Jarzynski's equality in the expanding relativistic piston model}

Jarzynski's equality, as a result of Liouville's theorem, should be directly generalized to any Hamiltonian systems.
In Appendix~\ref{sec:JE} we give a proof of the JE in a relativistic system.
It is worth emphasizing that the model we consider here is not characterized by a time-dependent Hamiltonian, but by a parametrized boundary condition \cite{Gong2016PRL}.
Thus the proof of the JE in Appendix~\ref{sec:JE} is inapplicable to the expanding rigid piston system.
Still, we can demonstrate that the JE is valid in the expanding rigid piston system.
In order to do so, we focus on the Jacobian determinant between the initial and the final state and show it to be unity.

Unlike the case of the low-speed limit, where the initial state satisfies the classical Maxwellian distribution, now the initial state distribution (Maxwell-J\"uttner distribution \cite{jut11}) at the temperature $T = 1/(k_B \beta)$ is
\begin{equation}
    f(x, p) = \frac{1}{2K_1(\frac{mc^2}{k_B T})mcL} e^{-\frac{mc^2}{k_B T}\sqrt{1+(\frac{p}{mc})^2}},
\end{equation}
where $K_1$ is the modified Bessel function of the second kind, $(x, p)$ are the initial position and the initial momentum, and $m$ is the static mass of the particle.
This distribution can also be expressed in the position-velocity space, with the initial velocity denoted by $v$, as
\begin{equation}\label{eqn:xvdistribution}
    F(x, v) = \frac{1}{2K_1(\frac{mc^2}{k_B T})cL} e^{-\frac{mc^2}{k_B T}\left(1-\frac{v^2}{c^2}\right) ^{-\frac{1}{2}}}\left(1 - \frac{v^2}{c^2}\right)^{-\frac{3}{2}}.
\end{equation}

The exponential work can be averaged over the initial distribution
\begin{equation}
    \langle e^{W/k_{B}T} \rangle = \frac{\int_0^L \mathrm{d}x \int_{-\infty}^{\infty} \mathrm{d}p \, e^{-\frac{mc^{2}}{k_{B}T}\sqrt{1+\left( \frac{p}{mc}\right)^2}} e^{W_{\tau}/k_B T}}{2K_1(\frac{mc^{2}}{k_{B}T})mcL}.
\end{equation}
Here $W_{\tau}$  can be uniquely determined by the initial state $(x, p)$.

What can be easily derived from our analysis is that a particle with an initial momentum $p$ can hit the moving piston for $n$ times and its momentum diminishes to

\begin{equation}\label{eqn:pn}
p_n = \frac{mv_{n}}{\sqrt{1-\left(\frac{v_n}{c} \right)^2}}.
\end{equation}

It is clear to see that the initial state, $(x, p)$, turns into the final state
\begin{equation}\label{eqn:xp}
    (x_{\tau}, p_{\tau}) = (|L - v_n \tau + (v_n + v_p)t_n|, p_n)
\end{equation}
at time $ \tau. $
With Eq.~(\ref{eqn:vn}), (\ref{eqn:tn}), and (\ref{eqn:pn}), the Jacobian determinant can be directly computed from Eq.~(\ref{eqn:xp}) (for details, see Appendix~\ref{app:der}),
\begin{equation}\label{eqn:Liou}
    \left|\frac{\partial(p_{\tau}, x_{\tau})}{\partial(p, x)}\right| = 1,
\end{equation}
and thus the JE can be verified in this expanding piston model.

\subsection{\label{subsec:workdistri}Distribution of work}

There are three dimensionless parameters in our model: $ \beta mc^{2}, $ $ L/c\tau $ and $ v_{p}/c. $
Therefore it is convenient to set $ m = c = k_{B} = L = 1, $ leaving only $ v_{p}, $ $ \tau $ and $ \beta $ as free parameters.

Using the probability distribution (\ref{eqn:xvdistribution}), we can evaluate the distribution of work $W$

\begin{equation}\label{eqn:pworiginal}
    P(W) = \int_{-1}^{1}\mathrm{d}x\int_{0}^{1}\mathrm{d}v\frac{e^{-\frac{\beta}{\sqrt{1-v^2}}}\delta(W - W_{\tau}(x, v))}{2K_1(\beta)(1 - v^2)^{\frac{3}{2}}},
\end{equation}
where $ W_{\tau} = 1/\sqrt{1-v^2} - 1/\sqrt{1-v_n^2}$ is the work done by the particles that have experienced $n$ collisions.

After some tedious calculations (see Appendix~\ref{app:int} for details), the distribution function of $W$ can be analytically expressed as

\begin{equation}\label{eqn:pw}
    \begin{split}
         P(W) = &P_0 \delta (W) + \frac{1}{2 K_1 (\beta) } \sum_{n = 1}^N \varphi_n (v_n (W)) \\
         \times &\frac{e^{- \beta/ \sqrt{1 - v_n (W)^2}}}{(\alpha_p^{- n} - 1) \left[ 1 + \alpha_p^n - v_n (W) (1 - \alpha_p^n) \right]} .
    \end{split}
\end{equation}
Here the overlap factor $\varphi_{n}(v)$ is a trapezoid-shaped function
    
    \begin{equation}\label{eqn:overlap}
         \varphi_n (v) =
         \begin{cases}
            1 - \xi_{n}(v), & \frac{X_{n} - 1}{T_{n}} < v_{n}(W) \leq \frac{X_{n} + 1}{T_{n}} \\
            2, &  \frac{X_{n} + 1}{T_{n}} < v_{n}(W) \leq \frac{X_{n + 1} - 1}{T_{n + 1}} \\
            1 + \xi_{n + 1} (v), & \frac{X_{n + 1} - 1}{T_{n + 1}} < v_{n}(W) \leq \frac{X_{n + 1} + 1}{T_{n + 1}}
        \end{cases},
    \end{equation}
where $v_{n}$ as an inverse function of $W_{\tau}$ can be expressed as
    \begin{equation}
    \begin{split}
        &v_n (W) = \\
        &\frac{(1 - \alpha_p^n)^3 (1 + \alpha^n_p) +
        4 W \sqrt{\alpha_p^{3 n} \left( (1 - \alpha_p^n)^2 + \alpha_p^n W^2 \right)}}{(1 - \alpha_p^{2 n})^2 + 4 \alpha_p^{2 n} W^2},
    \end{split}
    \end{equation}
and
\begin{equation}\label{eqn:criticalxv1}
    \xi_n(v) = -T_{n} v + X_{n},
\end{equation}
with
\begin{equation}
    T_n = \frac{ \alpha_p^{- (n - 1)} - 1 - \alpha_p + \alpha_p^n }{1 - \alpha_p} + \frac{\alpha_p^{- (n - 1)} + \alpha_p^n}{1 + \alpha_p} \tau,
\end{equation}
\begin{equation}
    X_n = \frac{ \alpha_p^{- (n - 1)} - \alpha_p^n }{1 - \alpha_p} + \frac{ \alpha_p^{- (n - 1)} - \alpha_p^n}{1 + \alpha_p} \tau .
\end{equation}
Here, $\xi_n(v)$ is a function of $v$ and denotes the initial position of the particles that happen to collide with the moving piston exactly $n$ times within time $\tau$ with the initial velocity $v$.
For a pictorial explanation, see Appendix~\ref{app:int}.
Note that $\xi_n(v)$ is a linear function of $v$. 
This is not as obvious as in the nonrelativistic regime.
For an intuitive explanation, see Appendix~\ref{app:cov}.

Just like the case in Newtonian mechanics, we expect the work distribution to have a Dirac $\delta$ peak at $W=0$.
Its amplitude can be simply evaluated as

\begin{equation}
    P_0 = \int_{0}^{1}\mathrm{d}v \frac{\varphi_{0}(v)e^{-\frac{\beta}{\sqrt{1-v^2}}}}{2K_1(\beta)(1 - v^2)^{\frac{3}{2}}},
\end{equation}
where the overlap function $\varphi_0(v)$ is piecewise linear.
It can be evaluated analytically, although it is quite involved.

\subsection{Nonrelativistic Limit}

In the nonrelativistic limit, we expect our results to recover those in Ref.~\cite{lua05}.
Here it is convenient to restore the speed of light $c$, and the nonrelativistic limit thus corresponds to $c$ approaching infinity.

To restore $c$ one needs to do some dimensional analysis.
With the presence of $c$, the dimensions of the relevant physical quantities are $[W] = \mathbf{T}^{-2}$, $[v_{n}(W)] = \mathbf{T}^{-1}$, $[P(W)] = \mathbf{T}^{2}$, $[\beta] = \mathbf{T}^{2}$, $[\tau] = \mathbf{T}$, and $[c] = \mathbf{T}^{-1}$, where $\mathbf{T}$ denotes the dimension of time.
We see that to restore $c$ one only needs to make the substitutions $W\mapsto W/c^2,$ $v_n(W) \mapsto v_n(W)/c,$ $\beta \mapsto \beta c^2,$ and $P(W) \mapsto P(W)c^2.$
We then have
\begin{equation}
    \begin{split}
         P(W) = &\frac{P_0 \delta (W/c^2)}{c^2} + \frac{1}{2 K_1 (\beta c^2) c^2} \sum_{n = 1}^N \varphi_n \left(\frac{v_n (W/c^2)}{c}\right) \\
         \times &\frac{e^{- (\beta c^2)/ \sqrt{1 - v_n (W/c^2)^2/c^2}}}{(\alpha_p^{- n} - 1) \left[ 1 + \alpha_p^n - v_n (W/c^2) (1 - \alpha_p^n)/c \right]} .
    \end{split}
\end{equation}

We will deal with the exponential part in Eq.~(\ref{eqn:pw}) separately.
For now we can expand the rest part of the work distribution except the exponential and the normalization constant at $ c $ around infinity.
We have
\begin{equation}
    \begin{split}
    &\frac{1 - \xi_n \left(v_n (W/c^2)/c\right)}{(\alpha_p^{- n} - 1) \left[ 1 + \alpha_p^n - v_n (W/c^2) (1 - \alpha_p^n)/c \right]} \\
    &= -\frac{(n - 1)c}{2nv_p}-\frac{(n-1)c\tau}{4n} +\frac{c\tau W}{8n^2v_p^2} + O(1),\\
    &\,\\ 
    &\,\\
    \end{split}
\end{equation}
\begin{equation}
\begin{split}
        &\frac{2}{(\alpha_p^{- n} - 1) \left[ 1 + \alpha_p^n - v_n (W/c^2) (1 - \alpha_p^n)/c \right]} \\
        &= \frac{c}{2nv_p} + O(1),
\end{split}
\end{equation}
and
\begin{equation}
    \begin{split}
    &\frac{1 + \xi_{n+1} \left(v_n (W/c^2)/c\right)}{(\alpha_p^{- n} - 1) \left[ 1 + \alpha_p^n - v_n (W/c^2) (1 - \alpha_p^n)/c \right]} \\
    &= \frac{(n + 1)c}{2nv_p}+\frac{(n+1)c\tau}{4n} - \frac{c\tau W}{8n^2v_p^2} + O(1)
    \end{split}
\end{equation}
as all pieces of the trapezoid-shaped function $ \varphi_n. $

The exponential term in Eq.~(\ref{eqn:pw}), together with the normalization constant $ K_{1}(\beta) $, needs to be treated with extra care.
We start by taking the limit $ c \to \infty $ of the exponential
\begin{equation} 
\begin{split}
    &-\frac{\beta c^2}{\sqrt{1-v_{n}\left(W/c^2\right)^{2}/c^2}} \\
    &= -\beta c^2 - \frac{\beta}{2} \left(\frac{W}{2nv_p} + nv_p\right)^2 + O(c^{-2}),
\end{split}
\end{equation}
where a temperature dependent constant $ e^{-\beta c^2} $ appears as the leading term.
The limit of $c \to \infty$ also affects the constant factors, yielding
\begin{equation}\label{eqn:expansionOfBesselK} 
\frac{e^{-\beta c^2}}{K_{1}(\beta c^2)} \approx \sqrt{\frac{2\beta}{\pi}} c,
\end{equation}
which is exactly the Maxwellian normalization constant multiplied by $c$.
Multiplying all pieces together we see that the $c$ dependence cancels, which agrees with the physical intuition that the low-speed physics is independent of the speed of light.

To conclude, we are able to recover the nonrelativistic results in Ref.~\cite{lua05} ($\tau = 1$ is chosen in accordance with Ref.~\cite{lua05})
\begin{equation}\label{eqn:luapw}
    P (W) = \frac{\sqrt{\beta}}{\sqrt{2\pi}nv_p} e^{- \frac{\beta}{2} (\frac{W}{2nv_p} + nv_p)^2 } f(W),
\end{equation}
with
\begin{widetext}
    \begin{equation}
        f(W) = 
        \begin{cases}
            -(n - 1)(1+\frac{v_{p}}{2}) +\frac{W}{4nv_p}, & (n - 1)(v_{p} + 2) < \frac{W}{2nv_{p}} \leq (n - 1)(v_{p} + 2) + 2 \\
            1, & (n - 1)(v_{p} + 2) + 2 < \frac{W}{2nv_{p}} \leq (n - 1)(v_{p} + 2) + 2 + 2v_{p} \\
            (n + 1)(1+\frac{v_{p}}{2}) -\frac{W}{4nv_p}, & (n - 1)(v_{p} + 2) + 2 + 2v_{p} < \frac{W}{2nv_{p}} \leq (n + 1)(v_{p} + 2)
        \end{cases}
\end{equation}
\end{widetext}
as a low-speed limit.

In Fig.~\ref{fig:figure1}-\ref{fig:figure3} we plot the work distributions of the expanding relativistic piston model and its nonrelativistic limit.
The deviations of the relativistic results from the nonrelativistic ones with different choices of parameters are shown.
One can see that as expected, the relativistic results deviate more prominently from the nonrelativistic results at high temperature and fast speed.
\captionsetup[figure]{justification=raggedright}
\begin{figure}[ht]
    \centering
    \begin{subfigure}{0.45\textwidth}
        \includegraphics[width=\linewidth]{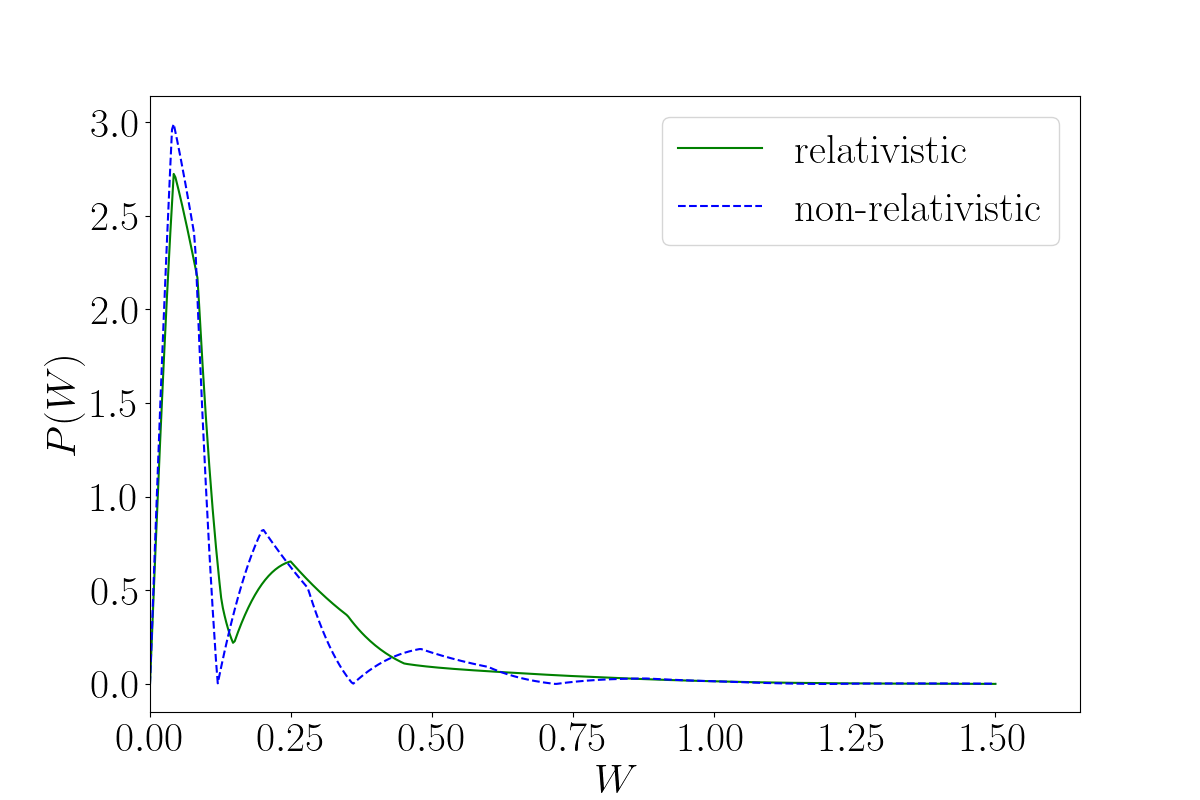}
        \caption{\label{fig:figure1}}
    \end{subfigure}
\end{figure}
\begin{figure}[H]
    \ContinuedFloat
    \centering
    \begin{subfigure}{0.45\textwidth}
        \includegraphics[width=\linewidth]{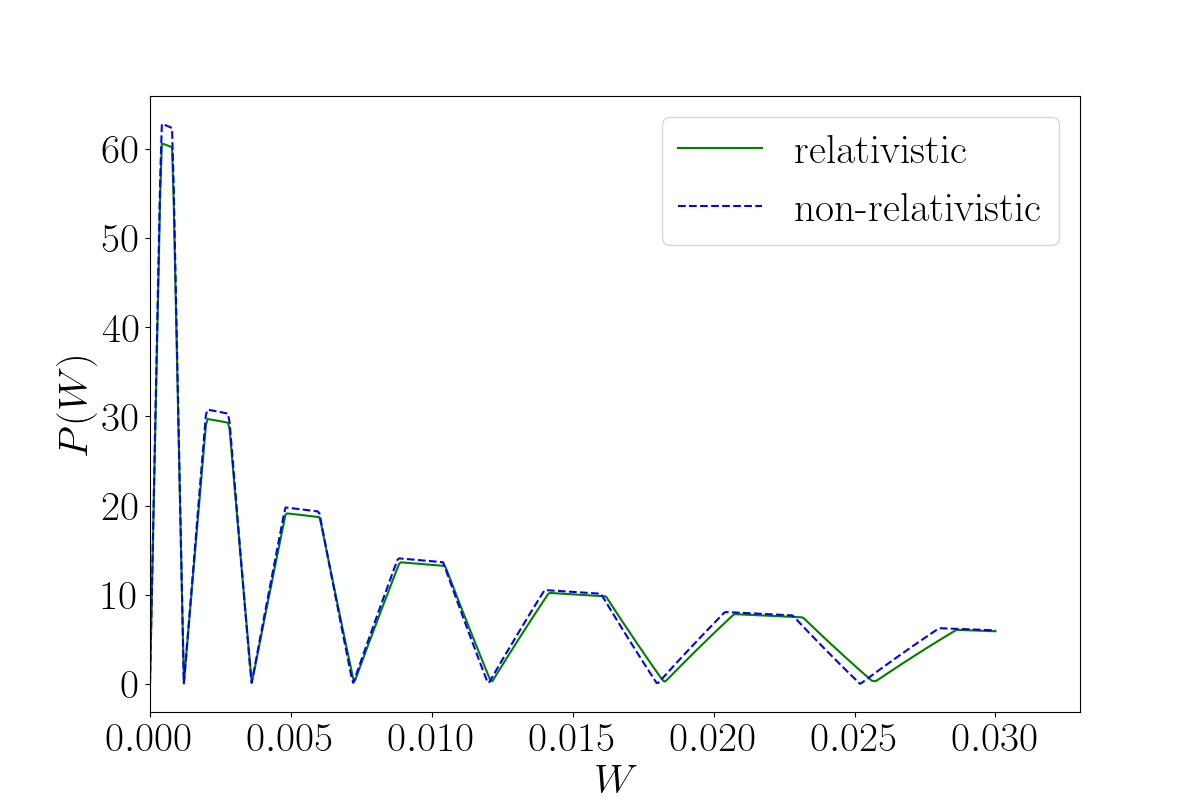}
        \caption{\label{fig:figure2}}
    \end{subfigure}
\end{figure}
\begin{figure}[H]
    \ContinuedFloat
    \centering
    \begin{subfigure}{0.45\textwidth}
        \includegraphics[width=\linewidth]{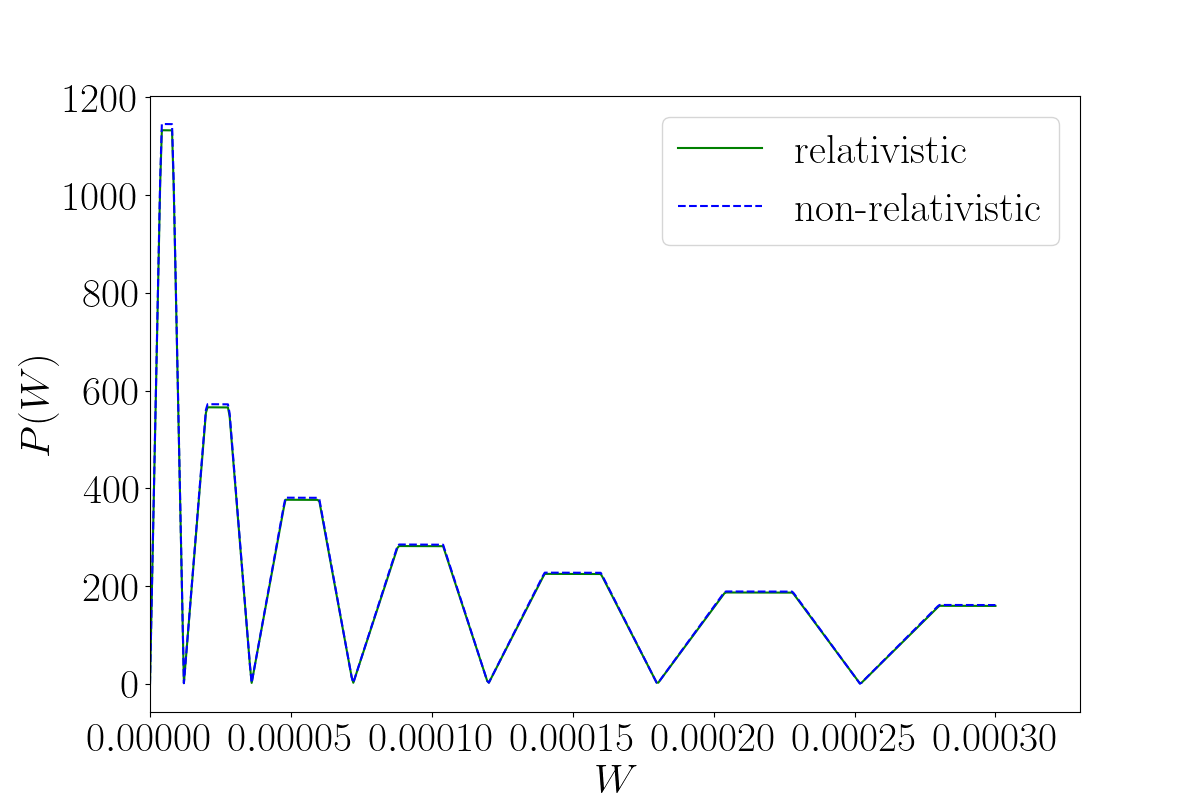}
        \caption{\label{fig:figure3}}
    \end{subfigure}
    \caption{Relativistic and nonrelativistic work distribution with different parameters. We use hydrogen atoms as an example and suppose that the length of the cylinder is $L= 1~\mathrm{cm} $ (which is much larger than the thermal length of the atoms). The parameters are chosen to be (a) $\tau=0.3~\mathrm{ns}, v_p=3\times 10^7~\mathrm{m/s}, T=3\times 10^{12}~\mathrm{K}$, and $\beta = 3$; (b) $\tau=3 ~\mathrm{ns}, v_p=3\times 10^6~ \mathrm{m/s}, T=1\times 10^{12}~\mathrm{K}$, and $\beta = 10$; and (c) $\tau=30~ \mathrm{ns}, v_p=3\times 10^5 ~\mathrm{m/s}, T=3\times 10^{11}~\mathrm{K}$, and $\beta = 33$}
\end{figure}

\subsection{Limit of sast-moving piston}
We already know that in Newtonian mechanics, at a very large $v_p$, the validity of the JE relies on the far tails of the Maxwellian distribution \cite{lua05}.
It is thus intriguing to also think of this problem in special relativity, where every speed has the speed of light $ c $ as its upper bound.
The main obstacle to the application of the JE is that to measure the average of exponential work, one must repeat the experiment a certain number of times.
The exponent makes sure that the contribution of the tail of the distribution, while its probability goes to zero, is nonvanishing \cite{Jar06PRE}.
Specifically in our model, when the speed of the moving piston approaches the speed of light, the fraction of particles that can collide with the piston approaches 0.
In such a case an experiment with nonzero work is of probability
\begin{equation}
    P(W>0)=\int_{v_p}^{1} \mathrm{d}v \frac{\varphi_1(v) e^{-\frac{\beta}{\sqrt{1-v^2}}}}{2K_1(\beta)(1-v^2)^{\frac{3}{2}}} .
\end{equation}

The expectation value of the exponential work is 
\begin{equation}\label{eqn:expwLowT}
    \begin{split}
        \left\langle e^{\beta W} \right\rangle 
        &= P_0 e^{\beta \cdot 0} \\
        &+\int_{v_p}^{1} \mathrm{d}v \frac{\varphi_1(v) e^{-\frac{\beta }{\sqrt{1-v^2}}}}{2K_1(\beta)(1-v^2)^{\frac{3}{2}}} e^{\beta W}.
\end{split}
\end{equation}
The first term is approximately equal to 1 ($P(W>0)\ll1$ and $P_0\approx 1$).
From the value of $\exp{(-\beta \Delta F)}=1+v_p\tau$, one can expect that, although the probability $P(W>0)$ is vanishingly small, the contribution of the second term is nonzero.
This result can be demonstrated transparently in the low-temperature limit.
Noticing that 
\begin{equation}
\begin{split}
    e^{-\beta (1-v^2)^{-\frac{1}{2}}} e^{\beta W}
    =&e^{-\beta H_{\lambda(0)}} e^{\beta (H_{\lambda(0)}-H_{\lambda(\tau)})} \\
    =&e^{-\beta H_{\lambda(\tau)}} \\
    =&e^{-\beta (1-v_1^2)^{-\frac{1}{2}}}, 
\end{split}    
\end{equation}
with $v_1$ derived from Eq.~(\ref{eqn:vn}), the particles with small $v_1$ after one collision contribute the most to the exponential work. 
When $v_1 = 0$, the corresponding initial velocity is $2v_p/(1+v_p^2)$.

When $\beta \to \infty $, $e^{-\beta (1-v^2_1(v))^{-\frac{1}{2}}}\approx e^{-\beta (1+v^2_1(v)/2)}$, and this term becomes a Gaussian peak, which is a $\delta $ distribution around $v=2v_p/(1+v_p^2)$,
\begin{equation}
\begin{split}
    e^{-\beta (1-v^2_1(v))^{-\frac{1}{2}}} 
    \approx &e^{-\beta} e^{-\frac{\beta}{2}v^2_1(v)}\\
    \approx &e^{-\beta}\sqrt{\frac{2\pi}{\beta}}\delta(v_1(v)) \\
    =&e^{-\beta}\sqrt{\frac{2\pi}{\beta}}\left[ 1-\left(\frac{2v_p}{1+v_p^2}\right)^2 \right]\delta \left(v-\frac{2v_p}{1+v_p^2}\right).
\end{split}
\end{equation}
Together with (\ref{eqn:expansionOfBesselK}) and (\ref{eqn:overlap}), the second term in Eq.~(\ref{eqn:expwLowT}) becomes
\begin{equation}
\begin{split}
    & \int_{v_p}^1 \mathrm{d}v \frac{e^{-\beta}\sqrt{\frac{2\pi}{\beta}}\left[ 1-\left(\frac{2v_p}{1+v_p^2}\right)^2 \right]\delta \left(v-\frac{2v_p}{1+v_p^2}\right)}{2 e^{-\beta} \sqrt{\frac{\pi}{2\beta}}} \frac{\varphi_1(v)}{(1-v^2)^{\frac{3}{2}}} \\
    =&\left[ 1-\left(\frac{2v_p}{1+v_p^2}\right)^2 \right]^{-\frac{1}{2}} \varphi_1\left(\frac{2v_p}{1+v_p^2}\right) \\
    =&v_p \tau .
\end{split} 
\end{equation}
This result demonstrates that when the piston moves at a very large $v_p$, particles with high initial velocities around $2v_p/(1+v_p^2)$ contribute most significantly to the exponential work even the probability is extremely small.
Note that in Newtonian mechanics particles with initial velocities around $2v_p$ contribute most significantly to the exponential work \cite{Quan2012PRE, Jar06PRE}, even the probability is extremely small.
It can be seen that the results of the relativistic piston model recover those of the nonrelativistic piston model as expected.

\section{Conclusion\label{sec:dis}}
Let us look further into the main results we have obtained.
We see that, as a consequence of the relativistic energyvelocity relation, the trapezoid-shaped work distribution no longer has a series of zeros\footnote{In fact, in the nonrelativistic regime, $ v_n(W) $ and $ v_{n+1}(W) $ never overlap, that is, there is only one collision number $ n $ corresponding to a particular value of $ W $. However, with the relativistic energyvelocity relation, a single value $ W $ may correspond to different collision numbers.}.
Moreover, the number of peaks becomes finite because the speed of light places an upper bound on all speeds.
The apparent paradox of the fast-moving piston in Ref.~\cite{lua05} can be reformulated as when the piston is moving at the speed of light instead of infinity.
No particle would be able to catch up with the piston and the average exponential work is unity whereas the free energy change is nonzero.
We would like to point out that although lightlike world lines exist, we cannot make it stop before and after the moving time period, because such a world-line configuration would violate causality.
The best we can do is to take the limiting process of letting the speed of the piston approach the speed of light.
Then the order of limit becomes crucial, as we have to integrate out the work $ W $ to infinity, and then take the speed limit \cite{Quan2012PRE}.
This limiting procedure ensures the validity of the JE.

In order to observe a distinct deviation of the relativistic work distribution from its nonrelativistic limit, the speed of the piston should be large enough and the temperature of the ideal gas should be extremely high.
Taking hydrogen atoms as an example, to observe the features of the relativistic work distribution, the piston should be as fast as about $3\times 10^6~\mathrm{m/s}$, and the temperature of the atoms should be about $10^{12}~\mathrm{K}$.
When the speed of the piston is about $3 \times 10^5~ \mathrm{m/s}$ (faster than the Parker Solar Probe, which is the fastest object human ever built) and the temperature of the atoms is $10^{11}~\mathrm{K}$ ($10^4$ times hotter than the central temperature of the Sun), the deviation of the relativistic work distribution from the nonrelativistic result becomes barely detectable.
Even in such a circumstance, the boundary condition is still difficult to realize.
Because the energy scale of the kinetic energy of the atoms is much larger than the energy scale of the chemical bond, the boundary cannot be built by any materials we have already discovered.
Based on these facts, we conclude that it is difficult to detect the relativistic effects of the work distribution of ideal gas in a piston system with the current experimental techniques.

In summary, we have studied a simple model of a piston and ideal gas in the framework of the special theory of relativity.
We obtained an analytical result of the work distribution (\ref{eqn:pw}) and verified the JE.
Using our result, it was possible to see the deviation of relativistic work distribution (\ref{eqn:pw}) from the nonrelativistic one (\ref{eqn:luapw}) \cite{lua05}.
In principle, these relativistic corrections become non-negligible in the high-temperature and fast-speed limit.
However, the range of parameters where relativistic effects are observable would already be far beyond the current experimental techniques.
Our results show that the JE holds true in a wide range of systems with generality and serve the pedagogical purpose.

\section*{Acknowledgments}
This work was supported by the National Natural Science Foundations of China under Grants No. 12375028 and No. 11825501.

\appendix

\section{Covariant description of the trajectory}\label{app:cov}
In Eq.~(\ref{eqn:recursionresult}), we extended the range of the initial position $x$ from $[0, L]$ to $[-L, L]$ in order to remove the negative sign related to the direction of $v$.
We also notice that in Eq.~(\ref{eqn:criticalxv1}), when $ t_n $ happens to be $ \tau $,  the critical initial position and the initial velocity satisfy a linear relation.
Here we introduce a coordinate transformation method to describe the trajectory of a particle, which will provide pictorial intuition of the two facts.

To begin with, let us consider the collision between the particle and the fixed boundary. 
In the reference frame of the fixed boundary (which is the same as the laboratory frame), the coordinate frame consists of an $x$ axis and $t$ axis, and the world line of the moving piston is $l$ (see Fig.~\ref{fig:wla}).
We could draw an auxiliary world line of the moving piston, named $l'$. The auxiliary world line $l'$ and the real world line $l$ are mirror-symmetric about the $t$ axis.
The world line of the particle is a polyline $I$-$A$-$B$. Point $I$ denotes the initial condition of the particle, point $A$ denotes the collision event between the particle and the fixed boundary, and point B denotes the next collision event between the particle and the moving piston.
Transformation of the particle's world line works as follows: the auxiliary event of event $B$ is point $B'$.
Points $B$ and $B'$ are mirror symmetric about the $t$ axis.
The auxiliary world line of the particle is $I$-$A$-$B'$.
The real world line's $AB$ part and the auxiliary world line's $AB'$ part are also mirror symmetric about the $t$ axis.
As a result of such transformation, the auxiliary world line of the particle $I$-$A$-$B'$ is a straight line, which eliminates the world line's direction change during a collision.
Meanwhile, in the frame of the fixed boundary, the auxiliary collision event $B'$ has the same time coordinate as event $B$. 
Therefore, we could use an auxiliary event to evaluate the time when a real collision takes place. 

We may also map the initial condition $I(x,-v)$ to an auxiliary initial condition $I'(-x,v)$, which is the same as the treatment for  Eq.~(\ref{eqn:recursionresult}). 
The real motion is that a particle starts from the initial position $I$ with a positive initial position $x$ but a negative initial velocity $-v$ (the particle moves away from the moving piston), and then the particle collides with the fixed boundary at event $A$.
Meanwhile, the auxiliary motion is that a particle starts from $I'$ with a positive initial velocity $v$ but a negative initial position $-x$. The auxiliary particle passes through the fixed boundary at event $A$ without any collision.
After event $A$, both the particle and the auxiliary particle move towards the moving piston, and finally collide with the piston at the same event $B$. 
The rest trajectories are equivalent, for both the particle with the initial condition $(x, -v)$ and the auxiliary particle with the initial condition $(-x, v)$. 
Because there is no work done during the collision event $A$, it is convenient to extend the range of $x$ from $[0, L]$ to $[-L, L]$ while limiting the range of $v$ from $[-c,c]$ to $[0, c]$ in the calculation of the work distribution.
\begin{figure}[ht]
    \centering
    \begin{subfigure}{0.4\textwidth}
        \centering
        \includegraphics[width=\textwidth]{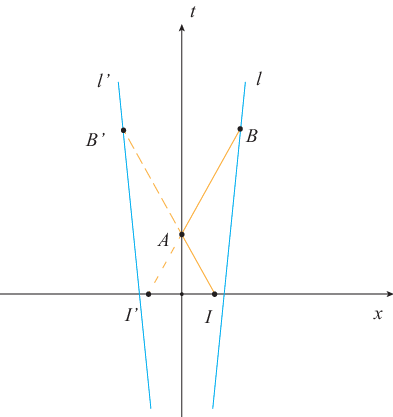}
        \caption{\label{fig:wla}}
    \end{subfigure}
\end{figure}
\begin{figure}
    \ContinuedFloat
    \centering
    \begin{subfigure}{0.4\textwidth}
        \centering
        \includegraphics[width=\textwidth]{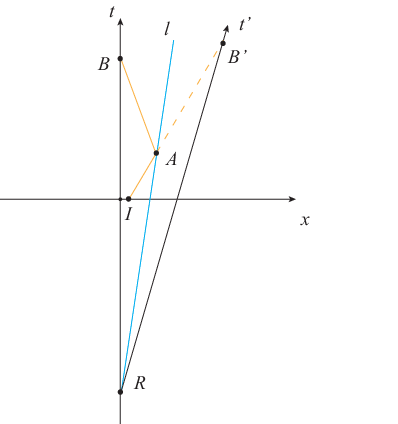}
        \caption{\label{fig:wlb}}
    \end{subfigure}
    \caption{Transformation of a world line. For details see the text Appendix~\ref{app:cov}.}
    \label{fig:transformworldline}
\end{figure}

After explaining the treatment in Eq.~(\ref{eqn:recursionresult}), we now continue to the understanding of Eq.~(\ref{eqn:criticalxv1}).
Similar to the mirror symmetric operation in the laboratory reference frame, we could deal with the collision between the particle and the moving piston by carrying out a mirror symmetric operation in the reference frame of the piston.
After a Lorentz transformation back to the laboratory reference frame, the result is shown in Fig.~\ref{fig:wlb}. 
Points $I$,$A$, and $B$ denote the initial state of the particle, a collision between the particle and the moving piston, and the next collision between the particle and the fixed boundary.
Point $B'$ is the auxiliary event of $B$.
Just like the case in Fig.~\ref{fig:wla}, in the reference frame of the moving piston, the auxiliary world line of the particle is a straight line.
Therefore, after a Lorentz transformation the auxiliary world line of the particle is still a straight line $I$-$A$-$B'$.
The auxiliary world line of the fixed boundary is denoted by $t'$. The velocity of $t'$ is $2 v_p / (1 + v_p^2)$, according to the Lorentz transformation.
The intercept of the line $t'$ can be determined as follows. The world line of the fixed boundary and moving piston intersect at point $R$ with space and time coordinate $(0,- L/ v_p)$. In the reference frame of the moving piston, the auxiliary world line of the fixed boundary and the moving piston intersect at the same point $R$.
Therefore, in the laboratory reference frame, the auxiliary world line of the fixed boundary $t'$ also passes through $R$, and the intercept of $t'$ must be $- L/ v_p$.

\begin{figure}[ht]
    \centering
    \begin{subfigure}{0.4\textwidth}
            \centering
            \includegraphics[width=\textwidth]
            {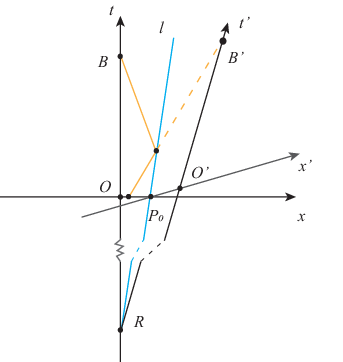}
            \caption{\label{fig:coordinatea}}
    \end{subfigure}
    \begin{subfigure}{0.5\textwidth}
            \centering
            \includegraphics[width=\textwidth]{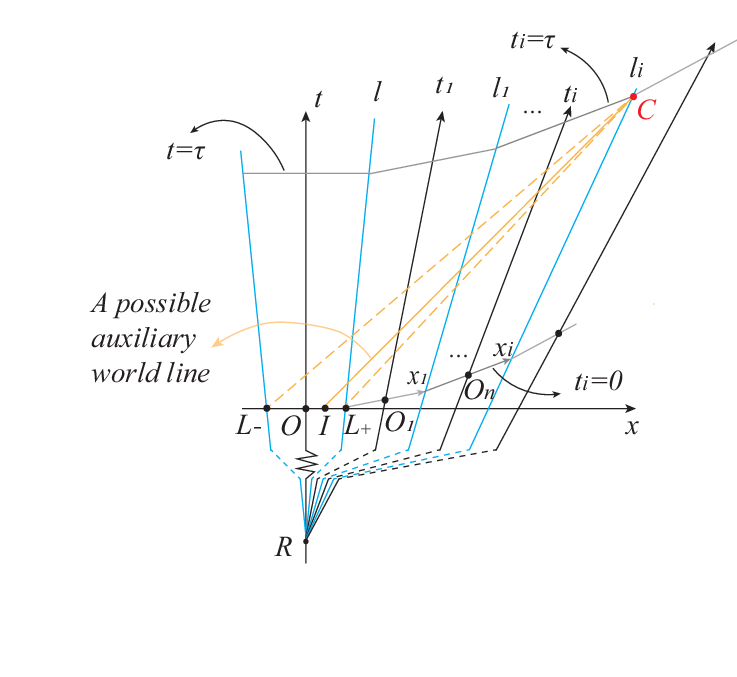}
            \caption{\label{fig:coordinateb}}
    \end{subfigure}
    \caption{Transformation of the coordinate frame and the complete world line. For details see the text Appendix~\ref{app:cov}.}
    \label{fig:transformcoordinate}
\end{figure}

The mirror-symmetric operation transforms not only the world lines, but also the coordinate frames.
In the reference frame of the fixed boundary, the $t$ axis is the world line of the boundary itself.
After a coordinate transformation related to a collision with the moving piston, the auxiliary world line of the fixed boundary represents the auxiliary $t'$ axis.
The auxiliary $x'$ axis should be orthogonal to the $t'$ axis, and the origin of coordinates (point $O$) should be transformed accordingly: The $x$ axis and $l$ intersect at $P_0$, and thus the $x'$ axis also passes through $P_0$ (see Fig.~\ref{fig:coordinatea}.
The intersection of the auxiliary $x'$ axis and the $t'$ axis is the auxiliary origin $O'$.
If we denote the time coordinate of event $B$ in the coordinate frame $xOt$ by $t_{B}$ and the time coordinate of event $B'$ in the coordinate frame $x'O't'$ by $t'_{B'}$, then, as a result of such coordinate transformation, $t'_{B'} = t_B$.
The method that we evaluate the time coordinate of a real event with the help of an auxiliary event is still valid when dealing with the collision with the moving piston.

By performing the coordinate transformation repeatedly, we could figure out the complete auxiliary world line of the particle and every auxiliary coordinate frame. 
The possible space-time region is $S_0 = \{ (x, t) | 0 \leq t \leq \tau, 0 \leq x \leq L+ v_p t \}$, and the auxiliary regions can be determined accordingly: Each gray line segment represents $t_i = 0$ or $t_i = \tau$, where $t_i$ is the time coordinate of an auxiliary event in the $i$th auxiliary coordinate frame.
In Fig.~\ref{fig:coordinateb}, regions between gray line segments are possible auxiliary regions. The red dot $C$ represents the event that a particle happens to collide exactly $i+1$ times when $t = \tau$. The position and time 
coordinate of C in the initial coordinate frame $xOt$ is denoted by $(x_C, t_C)$.
Every auxiliary world line $IC$ with the initial condition $I(x_I, v_I)$ that passes through $C$ satisfies the critical condition 
\begin{equation}
    x_I = x_C - v_I t_C
\end{equation}
which is the same as Eq.~(\ref{eqn:criticalxv1}) when $n=i+1$. Such world lines lie between the two orange dashed lines $L_- C$ and $L_+ C$ where $L_-$ and $L_+$ denote the particles with the initial positions $-L$ and $L$, respectively.
It is clear that the initial positions and velocities of auxiliary world lines satisfy a linear relation if every world line passes through a fixed point $C$, which explains the pictorial intuition mentioned above.

\section{Jarzynski's equality under the framework of relativistic mechanics\label{sec:JE}}

The original proof of the JE was based mainly on the classical Hamilton mechanics and its corollary, i.e., Liouville's theorem~\cite{jar97}.
In this Appendix, we extend the proof to the framework of relativistic mechanics.
We take the spacetime dimension to be $ 1+1 $ to suit our model.
Note that the generalization to higher spacetime dimensions is straightforward.
The manifestly covariant dynamics for a particle of the static mass $m$ can be formulated as
\begin{equation}
    f^{\mu} = m \frac{\mathrm{d} u^{\mu}}{\mathrm{d} \zeta},
\end{equation}
where $f^{\mu},\,u^{\mu},\,\zeta$ denote the 2-force, 2-velocity and proper time respectively.
To be clear, we fix our frame of reference to the laboratory frame, where the relativistic dynamics can be expressed more conveniently in the form of a 1-vector,
\begin{equation}\label{eqn:rellag}
    -\frac{\partial V}{\partial q} = \frac{\mathrm{d}}{\mathrm{d}t} \left( \frac{mu}{\sqrt{1-u^2/c^2}} \right),
\end{equation}
where $ V $ denotes the potential, $ c $ is the speed of light, and $q, u,$ and $t$ are the position, velocity and time, respectively measured in the laboratory frame.
With this form of dynamics, we can easily construct the Hamiltonian of a system of $N$ particles of mass $m_1,\,m_2,\,...,\,m_N$.
Furthermore, we let the Hamiltonian be controlled by an external agent via the parameter $ \lambda = \lambda(t) $ of the potential $V_{\lambda}.$
The Hamiltonian reads
\begin{equation}
    H_{\lambda} = \sum_{i = 1}^{N}\sqrt{m_i^2 c^4 + p_i^2 c^2} + V_{\lambda}(q_1,\,q_2,\,...,\,q_N),
\end{equation}
where $ q_i $ is the position of the $i$th particle and $ p_i $ its conjugate momentum.

The canonical equations are
\begin{equation}
    \frac{\mathrm{d}q_i}{\mathrm{d}t} = \frac{\partial H_{\lambda}}{\partial p_i} = \frac{p_i c^2}{\sqrt{m_i^2 c^4 + p_i^2 c^2}},
\end{equation}
\begin{equation}
    \frac{\mathrm{d}p_i}{\mathrm{d}t} = -\frac{\partial H_{\lambda}}{\partial q_i} = -\frac{\partial V_{\lambda}}{\partial q_i},
\end{equation}
which is exactly the relativistic velocity-momentum relation that reproduces Eq.~(\ref{eqn:rellag}).

As long as the canonical equations are formulated, one can easily generalize Liouville's theorem to relativistic regime~\cite{gol80}.
The theorem states that the Jacobian determinant of the canonical coordinates $(p_{i,t}, q_{i,t})$ at time $t$ as functions of the initial canonical coordinates $(p_{i,0}, q_{i,0})$ at time $0$ is
\begin{equation} 
\left| \frac{\partial(p_{i,t}, q_{i,t})}{\partial (p_{i,0}, q_{i,0})} \right| = 1.
\end{equation}

The initial equilibrium state of the inverse temperature $\beta$ is determined by a probability distribution $\rho(p_{i,0}, q_{i,0}) = \exp{[-\beta H(p_{i,0}, q_{i,0})]} / Z_{0}$ with $Z_{0}=\int \mathrm{d}p_{i,0}\mathrm{d}q_{i,0} \exp[-\beta H_{\lambda(0)}(p_{i,0}, q_{i,0})]$ the initial partition function.
We saw in Subsec.~\ref{subsec:verification}) that for ideal gas the distribution is the so-called Maxwell-J\"uttner distribution \cite{jut11}.
The proof of the JE follows by calculating the expectation value of the exponential work $ \exp{[W(p_i, q_i, \tau)]} $ done by the system along the trajectory up to time $\tau .$
By definition we have
\begin{equation}
    W(p_{i,0}, q_{i,0}, \tau) = H_{\lambda(0)}(p_{i,0},q_{i,0}) - H_{\lambda(\tau)}(p_{i,\tau},q_{i,\tau}) ).
\end{equation}
Note that our definition is different from the usual one by a minus sign.
So
\begin{equation}
    \begin{split}
        &\left\langle e^{\beta W} \right\rangle \\
        &= \int \mathrm{d}p_{i,0}\mathrm{d}q_{i,0}\,
        \rho(p_{i,0}, q_{i,0}) e^{\beta W(p_{i,0}, q_{i,0}, \tau)} \\
        &= \frac{1}{Z_{0}} \int \mathrm{d}p_{i,0}\mathrm{d}q_{i,0}\, e^{-\beta H_{\lambda(\tau)}(p_{i,\tau}, q_{i,\tau})}  \\
        &=\frac{1}{Z_{0}} \int \mathrm{d}p_{i,\tau}\mathrm{d}q_{i,\tau}\, e^{-\beta H_{\lambda(\tau)}(p_{i,\tau}, q_{i,\tau}))}  \\
        &= \frac{Z_{\tau}}{Z_{0}},
\end{split}
\end{equation}
where Liouville's theorem is used for the third equality and $Z_{\tau}$ denotes the final equilibrium state partition function.
The partition functions can be expressed in the form of free energy (see, for example, \cite{pat72}), resulting in
\begin{equation}
    \left\langle e^{\beta W} \right\rangle =  e^{-\beta (F_\tau - F_0)},
\end{equation}
where $F_\tau$ and $F_0$ are the free energies of the equilibrium states corresponding to $\lambda(\tau)$ and $\lambda(0)$.
Thus, we demonstrate the validity of the JE under the framework of relativistic mechanics.

\section{Derivation of Liouville's Theorem in the Relativistic Piston Model}\label{app:der}
Here we explain the derivation of Eq.~(\ref{eqn:Liou}) from Eq.~(\ref{eqn:xp}).
The Jacobian determinat is

\begin{equation}
\begin{split}
    \left|\frac{\partial(p_{\tau}, x_{\tau})}{\partial(p, x)}\right| 
    &=\left| \frac{\partial x_{\tau}}{\partial x} \frac{\partial p_{\tau}}{\partial p} - \frac{\partial x_{\tau}}{\partial p} \frac{\partial p_{\tau}}{\partial x} \right| \\
    &=\left| \frac{\partial x_{\tau}}{\partial x} \frac{\partial p_{\tau}}{\partial p} - \frac{\partial x_{\tau}}{\partial p}  0 \right| \\
    &= (v_{n} + v_{p})\frac{\partial t_{n}}{\partial x}  \frac{\mathrm{d} p_n}{\mathrm{d} p},
\end{split}
\end{equation}
where 
\begin{equation}
\begin{split}    
    \frac{\mathrm{d} p_{n}}{\mathrm{d} p} &= \frac{\mathrm{d} p_{n}}{\mathrm{d} v_{n}}  \frac{\mathrm{d} v_{n}}{\mathrm{d} v} \left( \frac{\mathrm{d} p}{\mathrm{d} v} \right)^{-1} \\
    &= \left(\frac{1- \frac{v^2}{c^2}}{1- \frac{v_{n}^2}{c^2}}\right)^{\frac{3}{2}} \frac{4 \alpha_{p}^{2n} c^{2}}{(c-v+\alpha_{p}^{2n}(c+v))^2} \\
    &= \frac{\alpha_{p}^{-n}\left(c-v+\alpha_{p}^{2n}(c+v)\right)}{2c}.
\end{split}
\end{equation}
Noting that $v_p = (1-\alpha_{p})c/(1+\alpha_{p}),$ we finally get
\begin{equation}
\begin{split}
    \left|\frac{\partial(p_{\tau}, x_{\tau})}{\partial(p, x)}\right| 
    &= \left(v_{n} + \frac{1-\alpha_{p}}{1+\alpha_{p}} c\right) \frac{\partial t_{n}}{\partial x}\\
    & \times\frac{\alpha_{p}^{-n}\left(c-v+\alpha_{p}^{2n}(c+v)\right)}{2c}\\
    &= \frac{2 \left((c+v) \alpha_{p}^{2 n}-\alpha_{p} (c-v)\right)}{(\alpha_{p}+1) \left((c+v) \alpha_{p}^{2 n}+c-v\right)}\\
    &\times \frac{(1+\alpha_{p}) \left(\alpha_{p}^n-\alpha_{p}^{n+1}\right)c}{(1-\alpha_{p}) \left((c+v) \alpha_{p}^{2 n}-(c-v)\alpha_{p}\right)}\\
    &\times \frac{\alpha_{p}^{-n}\left(c-v+\alpha_{p}^{2n}(c+v)\right)}{2c}\\
    &=1,
\end{split}
\end{equation}
which demonstrates that the Jacobian determinant is equal to unity.

\section{Details of the Integration}\label{app:int}
\begin{figure}[ht]
    \centering
    \includegraphics[width=\linewidth]{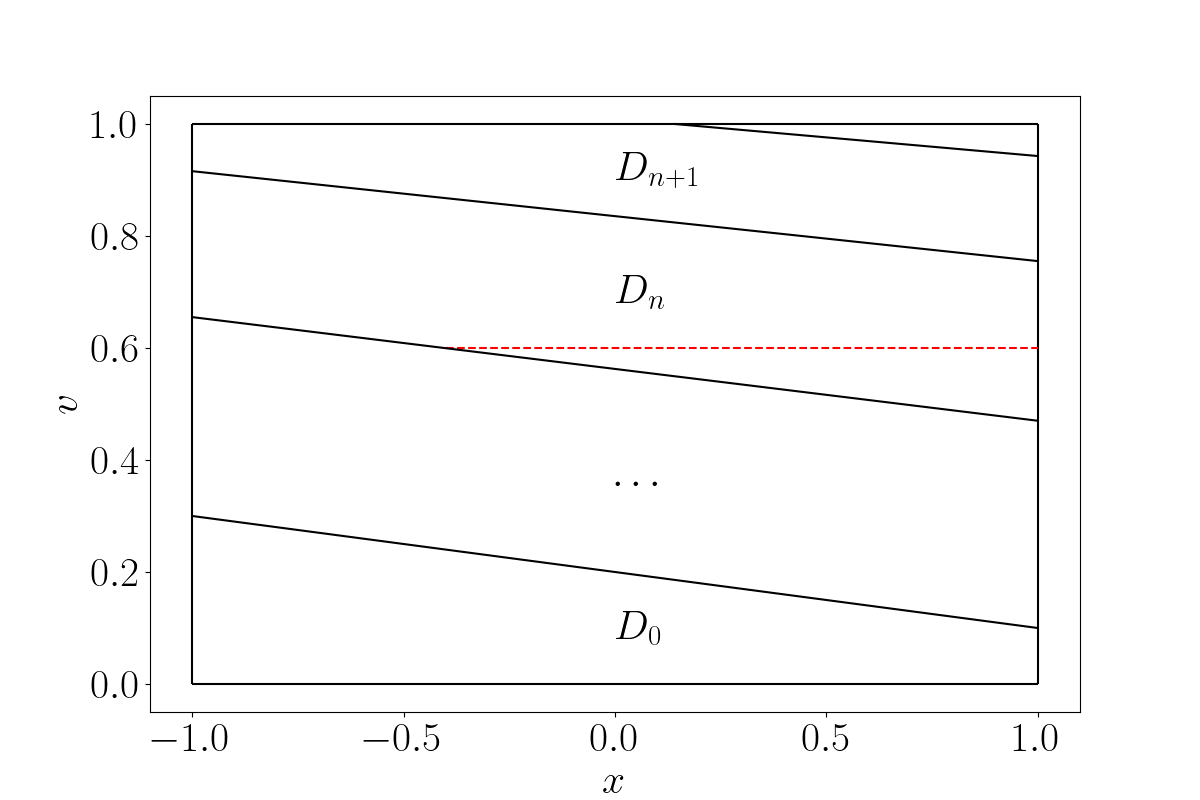}
    \caption{Division the domain of integration into parts. The overlap factor can be explained by the length of the red dashed line being the value of $\varphi_{n}$ at the same $v$.}
    \label{fig:reg}
\end{figure}

Here we explain the computational details of deriving Eq.~(\ref{eqn:pw}) from Eq.~(\ref{eqn:pworiginal}) and give a pictorial explanation for the overlap factor $\varphi_n.$ The integration

\begin{equation}
    P(W) = \int_{-1}^{1}\mathrm{d}x\int_{0}^{1}\mathrm{d}v\frac{e^{-\frac{\beta}{\sqrt{1-v^2}}}\delta(W - W_{\tau}(x, v))}{2K_1(\beta)(1 - v^2)^{\frac{3}{2}}}
\end{equation}
can be separated into parts
\begin{equation}
    P(W) = \sum_{n}\int_{D_{n}}\mathrm{d}x\mathrm{d}v\frac{e^{-\frac{\beta}{\sqrt{1-v^2}}}\delta(W - W_{\tau}(v))}{2K_1(\beta)(1 - v^2)^{\frac{3}{2}}},
\end{equation}
where $D_{n}$ is the domain of integration for all the $(x, v)$ values that the particle collide $n$ times.
Note that within each domain $D_{n},$ the trajectory work becomes independent of $x.$
Recall that those particles that collide exactly $n$ times lie in a straight line on the $x$-$v$ plane, with the equation
\begin{equation}\label{eqn:criticalxv}
    \xi_n(v) = -T_{n} v + X_{n},
\end{equation}
where
\begin{equation}
    T_n = \frac{ \alpha_p^{- (n - 1)} - 1 - \alpha_p + \alpha_p^n }{1 - \alpha_p} + \frac{\alpha_p^{- (n - 1)} + \alpha_p^n}{1 + \alpha_p} \tau
\end{equation}
and
\begin{equation}
    X_n = \frac{ \alpha_p^{- (n - 1)} - \alpha_p^n }{1 - \alpha_p} + \frac{ \alpha_p^{- (n - 1)} - \alpha_p^n}{1 + \alpha_p}  \tau
\end{equation}
are the slopes and the intercepts of the lines, respectively.
The separation of the domain of the integration is depicted in Fig.~\ref{fig:reg}.

We see that, since in each part the integrand is independent of $x$, we can integrate it out first, giving rise to
\begin{equation}
\begin{split}
     &\int_{D_{n}}\mathrm{d}x\mathrm{d}v\frac{e^{-\frac{\beta}{\sqrt{1-v^2}}}\delta(W - W_{\tau}(v))}{2K_1(\beta)(1 - v^2)^{\frac{3}{2}}} \\
     &= \int\mathrm{d}v\frac{e^{-\frac{\beta}{\sqrt{1-v^2}}}\delta(W - W_{\tau}(v))}{2K_1(\beta)(1 - v^2)^{\frac{3}{2}}} \varphi_{n}(v),
\end{split}
\end{equation}
where the value of $\varphi_{n}(v)$ is the length of the line segment shown in Fig.~\ref{fig:reg}. With the equation of all the lines known, we can compute the overlap factor as Eq.~(\ref{eqn:overlap}).

What is left involves integrating a Dirac $\delta$ function.
We have
\begin{equation}
    \begin{split}
        &\int\mathrm{d}v \frac{e^{-\frac{\beta}{\sqrt{1-v^2}}}\delta(W - W_{\tau}(v))}{2K_1(\beta)(1 - v^2)^{\frac{3}{2}}} \varphi_{n}(v) \\
        &= \left.\frac{e^{-\frac{\beta}{\sqrt{1-v^2}}}}{2K_1(\beta)(1 - v^2)^{\frac{3}{2}}} \varphi_{n}(v) \left(\frac{\mathrm{d}W_{\tau}}{\mathrm{d}v}\right)^{-1}\right\vert _{W_{\tau}(v) = W} \\
        &= \varphi_n (v_n (W)) \frac{ e^{ -\frac{\beta}{\sqrt{1 - v_n (W)^2}} } }{(\alpha_p^{- n} - 1) \left[ 1 + \alpha_p^n - v_n (W) (1 - \alpha_p^n) \right]}.
    \end{split}
\end{equation}

The $n = 0$ case should be treated separately. This is because all particles that cannot catch up with the piston contribute to the probability at $W = 0,$ resulting in a $\delta$ peak with the amplitude
\begin{equation}
    P_0 = \int_{0}^{1}\mathrm{d}v \frac{\varphi(v)e^{-\frac{\beta}{\sqrt{1-v^2}}}}{2K_1(\beta)(1 - v^2)^{\frac{3}{2}}}.
\end{equation}
Summing up all the pieces at hand we have the result (\ref{eqn:pw}).

\vspace{7em}

\bibliography{template}

\end{document}